\documentclass{mn2e}
\usepackage{graphicx}

\begin{document}

\title
  {Supernovae in isolated galaxies, in pairs and in groups of galaxies}
\author[H. Navasardyan et al.]
  {H.~Navasardyan$^1$\thanks{e-mail:  hripsime@bao.sci.am}, A.R.~Petrosian$^1$,
  M.~Turatto$^2$\thanks{e-mail: turatto@pd.astro.it}, E.~Cappellaro$^2$,
  J.~Boulesteix $^3$ \\
  $^1$Byurakan Astrophysical Observatory and Isaac Newton
      Institute of Chile, Armenian Branch,
      378433 Byurakan, Armenia\\
  $^2$Osservatorio Astronomico di Padova, 5 Vicolo dell'Osservatorio, I - 35122, Padova,
  Italy\\
  $^3$Observatoire Astronomique de Marseille - Provence au Laboratoire d'Astrophysique de
      Marseille, 2 Place Le Verrier I-13248, \\
      Marseille Cedex 4, France}

%\date{}
%\pubyear{2001}

\maketitle

\begin{abstract}
 In order to investigate the influence of the environment on the
supernova (SN) production we have performed a statistical
investigation of the SNe discovered in isolated
galaxies, in pairs and in groups of  galaxies.  22 SNe  in 18
isolated galaxies, 48 SNe in 40 galaxies members of 37 pairs and
211 SNe in 170 galaxies members of 116 groups have been
selected and studied.

We found that the radial distributions of core-collapse SNe in galaxies 
located in different environments are
similar, and consistent with that reported by Bartunov, Makarova \&
Tsvetkov (1992). SNe discovered in pairs do not privilege a
particular direction with respect to the companion galaxy. Also
the azimuthal distributions inside the hosts members of galaxy
groups are consistent with being isotropics. The
fact that SNe are more frequent in the brighter components of the
pairs and groups  is  expected from the dependence of the SN rates
on the galaxy luminosity.

There is an indication that the SN rate is higher in 
galaxy pairs compared with that in groups. This can be related to 
the enhanced star formation rate in strongly interacting systems.

It is concluded that, with the possible exception 
of strongly interacting system, the parent galaxy environment has no direct
influence on the SN production

\end{abstract}


\section{Introduction}

The stellar content and the history of the star formation (SF) are
the key parameters determining the evolution of galaxies. In this
respect the determination of the star formation rates (SFRs) in
the galaxies is a first step toward the characterization of
different systems.

Among the various methods conceived to estimate the SFRs in galaxies the
most widely used are based on the integrated colors, the emission lines,
the UV and FIR luminosities (e.g. Kennicutt 1998).

It is often claimed that different factors enhance the star
formation in galaxies above the average value for the given
morphological type. In particular, several authors have proposed
gravitational interaction as a possible SF triggering mechanism.
Evidences in this respect come from statistical studies showing
that the fraction of interacting galaxies is higher than average
in Markarian objects (e.g. Heidmann \& Kalloghlian 1973, Keel \&
van Soest 1992). Further the peculiar colors of interacting
systems are attributed to the occurrence of bursts of star
formation (e.g. Larson \& Tinsley 1978, Kennicutt et al. 1987).
Finally, {\it IRAS\/} observations have shown that interacting
systems emit a higher infrared luminosity than isolated ones which
can be combined with the knowledge that very luminous infrared
galaxies have experienced strong nuclear starburst (e.g., Soifer
et al. 1984, Sanders et al. 1988, Melnick \& Mirabel 1990, Wu et
al. 1998). Detailed investigations of the relation between
interaction and star formation show that most of star formation
takes place in the central regions of the galaxies  (e.g. Hodge
1975, Arp 1973, Laurikainen \& Moles 1989, Petrosian \& Turatto
1995, Laurikainen, Salo \&  Aparicio 1998) and in the tidal
streams (e.g. Schombert, Wallin \& Struck-Marcell  1990, Thronson
et al. 1989). Whether SF is stimulated also in less dense
environments such as compact groups of galaxies is still
questioned. For instance, Moles et al. (1994) find a slight
enhancement of SF in the galaxies members of compact groups, while
Iglesias-Paramo \& Vilchez (1999) do not confirm such finding.

It is now widely accepted that most SNe can be assigned to two
main physical classes, core collapse (type II and Ib/c SNe) and
thermonuclear explosions (type Ia SNe).

The rates per unit blue luminosity of type Ia SNe, were found to
increase (Oemler \& Tinsley 1979, Cappellaro et al. 1993b, 1997)
or at least to remain constant (Cappellaro, Evans \& Turatto 1999)
moving from early to late type galaxies. Because in late spirals
there is a significant contribution to the blue luminosity by
young stellar population this immediately implies that  the
average age of SNIa progenitors in spirals is shorter than in
ellipticals.

Kochhar (1989, 1990) has suggested that all SNIa are short-lived
stars, hence that early type SN host galaxies have on-going star
formation. However, Turatto, Cappellaro \& Benetti (1994) showing
that the rate of SNIa in ellipticals does not depend on the gas or
dust content, did not confirm this conclusion.

Type II SNe originate from young massive stars associated to the
ongoing process of star formation (e.g. Branch et al. 1991) and
this is why they appear concentrated in  star forming sites (e.g.
Maza \& van den Bergh 1976, Bartunov, Tsvetkov \& Filimonova 1994,
Van Dyk 1992, Van Dyk , Hamuy \& Filippenko 1996, Van Dyk et al.
1999). The discovery of intermediate cases (e.g. SN 1993J) sharing
the properties of SNII at early time and those of SNIb in the
nebular phase is the observational proof that also SNIb/c are also
related to massive progenitors.

Cappellaro et al. (1999) have shown that the rate of core-collapse SNe is related
to the FIR excess and to the color of galaxies, justifying the use of  the rate
of core collapse SNe as SF indicators.

Following the approach of Turatto, Cappellaro \& Petrosian (1989)
and Petrosian \& Turatto ( 1990, 1992, 1995), in this paper we
will use the rates and the locations of SNII and Ib/c in order to
study the relation between the star formation in galaxies and the
galaxy environment, in selected samples of  isolated galaxies,
pairs of galaxies and groups. The same study is performed also for
Ia type SNe.

The samples of SNe in isolated galaxies as well as in members of pairs and groups
of galaxies are presented in Section 2. The rates, radial and azimuthal distributions
in the parent systems along with the results of the Multivariate Factor Analysis are
analyzed and discussed in Section 3. Conclusions are drown in Section 4.

Through this article we have assumed for the Hubble constant a value of
$H_{0}$ = 75 km $s^{-1}$Mp$c^{-1}$.
%----------------------------------------------------------------------------------------------

\section{The samples}
The selection of the samples has been done by crossing a working version (updated to
mid 1998) of the Asiago Supernova Catalogue (Barbon et al.  1984, 1989, 1999, hereafter
ASC)  with:
\begin{enumerate}
 \item the catalogue of Isolated Galaxies ( Karachentseva 1973, hereafter KIG ),
 \item the Catalogues of isolated pairs of galaxies in the northern and southern hemispheres
(Karachentsev 1972, Reduzzi \& Rampazzo 1995 hereafter KPG , RRPG
respectively)
 \item the Catalogue of Nearby  Groups of Galaxies (Garcia 1993, hereafter LGG).
\end{enumerate}

It turns out that 22 SNe have been discovered in 18 isolated
galaxies. Table 1 summarizes the main data for these SNe and their
parent galaxies. In the first column is the galaxy running number
according to KIG and in column 2 is the most common identification
for the galaxy. In column 3 and 4, respectively, is the
morphological type and the activity class as reported in NED. When
for the same object alternative activity classifications are given
in the literature (Seyfert (Sy) or Liner (L)) we preferred the Sy
classification. Column 5 reports the absolute blue magnitude
($M_B$) of SN host galaxy from the Lyon-Meudon Extragalactic
Database (LEDA). Logarithmic ratios of the FIR to B luminosities
of SNe host galaxies (log($L_{FIR}/L_B$)) are listed in column 6
where the FIR luminosity was computed according to Lonsdale et al.
(1985). SN designations and classifications from ASC are presented
in columns 7 and 8, respectively. The relative distance of the SN
from the center of parent galaxy corrected for the tilting of the
galaxy along the line of sight ($R_0/R_{25}$, McCarthy 1973) is in
column 9. The last column 10 reports individual notes. It is worth
noting that because of the different criteria used, two galaxies
of the isolated sample belong also to the groups as defined by
Garcia (1993).

In isolated galaxies out of 22 SNe, 14 are type II+Ib/c, 4 are
type Ia and 4 are unclassified. 55\% of the SNe discovered in
spirals have barred hosts. 18\% of hosts have Sy nuclei which
become 22\%  if we include also Liners.

\setcounter{table}{0}
\begin{table*}
\centering
\begin{minipage}{140mm}
\begin{tiny}
\caption{Known SN hosts among isolated galaxies.}
\begin{tabular}{@{}lllcccllcl@{}}
 \hline
Name of & Galaxy  & Type &   AGN & $M_{B}$ & Log($L_{FIR}/L_{B}$)&
SN & & $R_{0}/R_{25}$& Notes\\
(1) & (2) & (3) & (4) & (5) & (6) & (7) & (8) & (9) & (10)\\\hline
KIG15  & CGCG 479-  4  &  Sc  &  &  -21.58 &  & 1954 F & & 0.52  \\
KIG104 & UGC  1903  & SBb  & & -21.31  &  & 1964 N & & 0.70 \\
KIG138 & UGC  2936 &  SBd   & & -20.33 &   0.04  &  1991 bd &Ia & 0.50 \\
KIG197 & NGC 2403  & SABcd   &  &  -19.57  & -1.35 & 1954 J & V & 0.27  \\
KIG309 & NGC 2775  & Sab  &  & -20.56 & -1.16  & 1993 Z &Ia & 0.40 & LGG169(2) \\
KIG324 & NGC  2841  & Sb  & Sy1 & -20.83 & -1.59  & 1912 A & & 0.34 \\
KIG324 & NGC  2841 &  Sb & Sy1 & -20.83 & -1.59  & 1957 A  &Ia & 0.69\\
KIG324 & NGC 2841  & Sb  & Sy1 & -20.83 & -1.59  & 1972 R & Ib & 0.78\\
KIG358 & NGC 2954 & E  &  &  -20.53 & & 1993 C & Ia & 0.99\\
KIG442 & NGC 3359  & SBc & & -20.75 & -1.23  & 1985 H & II & 0.24\\
KIG464 & NGC 3526 & Sc & & -19.02 & -1.05 & 1995 H  & II  & 0.54 \\
KIG469 & NGC 3556 & SBcd & & -20.99 & -0.95 & 1969 B  & II & 0.32 \\
KIG549 & NGC 4651 & Sc & L & -19.53 & -0.88 & 1987 K  & IIb & 0.20 \\
KIG605 & NGC 5375 & SBab & & -20.25 & -1.27 & 1989 K & II & 0.79\\
KIG604 & NGC 5377 & SBa  & & -20.25 & -1.17 & 1992 H & II & 0.56\\
KIG610 & NGC 5457 & SABcd & & -20.87 & -0.55 & 1909 A & II & 0.86 & LGG371(1)\\
KIG610 & NGC 5457 & SABcd & & -20.87 & -0.55 & 1951 H & II & 0.41 & LGG371(1)\\
KIG610 & NGC 5457 & SABcd & & -20.87 & -0.55 & 1970 G & II & 0.45 & LGG371(1) \\
KIG772 & IC 1231 & Scd & & -21.47 & -1.39 & 1976 C? & & 0.80\\
KIG812 & NGC 6389 & Sbc & & -21.22 & -0.91 & 1992 ab & II & 1.10\\
KIG967 & NGC 7292 & IBm & & -18.61 & -1.04 & 1964 H & II & 0.46\\
KIG1004 & NGC 7479 & SBc & Sy2 & -21.40 & -0.40 & 1990 U & Ic & 0.46\\
\hline

\end{tabular}
\end{tiny}
\end{minipage}
\end{table*}

Table 2 lists the 48 SNe discovered in 40 galaxies members of 37
pairs of galaxies. The first six columns of Table 2 contain the
same information as in Table 1 while Column 7 lists the
logarithmic ratio of the blue luminosity of the SN host galaxy to
that of its neighbor. Velocity differences of the members of the
pairs are given in Column 8.  In order to take only physical
systems we have excluded all pairs for which the velocity
difference between the components is higher than 500 km $s^{-1}$
or is unknown. Columns 9, 10 and 11 of Table 2 respectively are
the same with Columns of 7, 8, 9 of Table 1. In addition, Column
12 lists the ratio of the projected distance of SN from the center
of its host galaxy to the projected distance of the pair members.
Column 13 reports the position angle (PA) of the SNe with respect
to the line connecting the center of the parent galaxy to its
neighbor. The PAs are calculated clockwise with the neighbor
placed at $180^\circ$. Notes are presented in Column 14 .

In pairs of galaxies out of the 48 SNe 9 are type Ia, 19 are type
II+Ib/c and the remaining are either type I or unclassified. A bar
is present in the 65\% of the spiral parent galaxies. 
In this sample, 8\% of the SNe occurs in 
active nuclei hosts (19\% with Liners). SNe are discovered more
frequently in the luminous component of the pairs (72 \% of cases)
than in the fainter one.

\begin{table*}
\centering
\begin{minipage}{170mm}
\begin{tiny}
\caption{Known SN hosts among pair member galaxies.}
\begin{tabular}{llllccccllcccl}

 \hline
 Name     of   &  Galaxy & Type &   AGN   &  $M_{B}$  &
Log & Log & $\triangle V $ & SN  & & $\frac{R_{0}}{R_{25}}$ &
$\frac{R_{SN}}{R_{1,2}}$ & PA & Notes \\
& & & & & ($\frac{L_{FIR}}{L_{B}}$)& ($\frac{L_{P}}{L_{N}}$)\\
 (1)& (2) & (3) & (4) & (5) & (6) & (7) & (8) &
(9) & (10) & (11) & (12)  & (13) & (14)\\\hline
 KPG127A & NGC2276 & SABc    &   & -21.34 & -0.36  & 0.18  &  409 &
1962 Q & & 0.44 &   0.09   & 43\\
KPG127A & NGC2276 & SABc    &  &  -21.34 & -0.36 & 0.18 & 409 &
1968V &II & 0.63 & 0.13  &  339 \\
KPG127A & NGC2276 & SABc & & -21.34 & -0.36 & 0.18 &409 &1968W
& & 0.12 & 0.03 & 340\\
KPG127A & NGC2276 & SABc & &-21.34 & -0.36 & 0.18  &  409 & 1993X
& II & 0.89 & 0.20   & 271 \\
KPG132A & NGC2336 & SABbc & & -22.31 & -1.54 & 0.78 & 148 & 1987 L
& Ia & 0.62 & 0.08 & 202 \\
KPG150B & NGC2487 & SBb & & -21.32 & -0.95 & 0.34 & 383 & 1975 O &
Ia & 0.44 & 0.09 & 136 & LGG152(3)\\
KPG156A & NGC2535 & Sc & & -21.26 & -0.58 & 0.55 & 63 & 1901 A & &
0.27 & 0.19 & 270 \\
KPG175A & NGC2672 & E1-2 & & -21.40 & & 0.65 & 406 & 1938 B & &
0.41 & 1.08 & 126 \\
KPG210B & NGC2968 & I0 & & -19.20 & & -0.34 & 269 & 1970 L & I &
2.05 & 0.38 & 339 \\
KPG218A & NGC3031 & Sab & Sy1.8 & -21.56 & -2.56 & 1.20 & 349 &
1993 J & II b & 0.29 & 0.08 & 346 & LGG176(3) \\
KPG228B & NGC3169 & Sa & L & -20.47 & -0.77 & 0.03 & 65 & 1984 E &
II L & 0.68 & 0.14 & 146 & LGG192(4) \\
KPG234A & NGC3226 & E2 & L & -19.25 & & -0.37 & 177 & 1976 K & I &
0.37 & 0.25 & 284 & LGG194(8) \\
KPG234B & NGC3227 & SABc & Sy1.5 & -20.18 & -0.73 & 0.37 & 177&
1983 U & Ia & 0.16 & 0.15 & 249 & LGG194(9) \\
KPG281A & NGC3646 & Ring & &-22.72 & & 1.10 & 181 & 1989 N & II &
0.51 & 0.11 & 42 \\
KPG282A & NGC3656 & I & &-19.97 & -0.35 & 0.79 & 30 & 1973 C & &
0.45 & 1.18 & 110 \\
KPG282A & NGC3656 & I & &-19.97 & -0.35 & 0.79 & 30 & 1963 K & I &
0.26 & 0.67 & 212 \\
KPG288A & NGC3690 & SBm? & &-22.18 & 0.33 & 0.24 & 99 & 1990 al?&
& 0.33 & 0.86 & 338 \\
KPG288A & NGC3690 & SBm? & &-22.18 & 0.33 & 0.24 & 99 & 1992bu? &
& 0.12 & 0.26 & 284 \\
KPG288A & NGC3690 & SBm? & &-22.18 & 0.33 & 0.24 & 99 & 1993 G &
II & 0.31 & 0.66 & 219 \\
KPG332B & NGC4302 & Sc & &-20.02 & & 0.14 &11 & 1986 E & II &
0.71&0.85 & 113 & LGG289(6) \\
KPG334A & NGC4382 & S0 & &-20.43 & & 0.50 & 50 & 1960 R & Ia &
0.63 & 0.28 & 84 & LGG292(35) \\
KPG335B & NGC4410B & SABcd & & -21.16 & & -0.14 & 30 & 1965 A & I&
1.27 & 1.00 & 27 \\
KPG336B & NGC4411B & SABcd & & -18.50 & -1.09 & 0.16 & 22 & 1992
ad & II & 0.58 & 0.17 & 303 & LGG289(39) \\
KPG341B & NGC4490 & SBd & & -20.67&  -0.81 & 1.04 & 23 & 1982 F &
II & 0.4 & 0.18 & 49 & LGG290(8) \\
KPG345A & NGC4512 & SBdm & &-18.99 & -1.00 & -0.46 & 454 & 1995 J&
II & 0.79 & 0.17 & 150 & LGG295(2) \\
KPG347B & NGC4568 & Sbc & & -21.61 & & 0.30 & 47 & 1990 B & Ib/c &
0.08 & 0.15 & 219 & LGG285(24) \\
KPG348B & NGC4615 & Scd & & -21.10 & -0.61 & 0.38 & 78 & 1987 F &
IIn & 0.52 & 0.15 & 291 \\
KPG349A & NGC4618 & SBm & &-19.04 & -1.17 & 0.60 & 8 & 1985 F &
Ib/c & 0.06 & 0.03 & 143 & LGG290(10) \\
KPG353A & NGC4647 & SABc & & -19.93 & -0.56 &-0.62 & 307 & 1979 A&
I & 0.68 & 0.38 & 14 & LGG289(49) \\
KPG379A & NGC5194 & Sbc & Sy2 & -20.50 & -0.52 & 0.24 & 107 & 1994
I & Ic & 0.07 & 0.07 & 64 & LGG347(4) \\
KPG379B & NGC5195 &SB0 & L & -19.90 & -1.37 & -0.24 & 107 & 1945 A
& I & 0.04 & 0.03 & 140 & LGG347(8) \\
KPG411B & M+02-36-26 & Sb & & -21.58 & & -0.03 & 291 & 1951 B & &
0.52 & 0.45 & 200 &LGG372(3) \\
KPG416A & NGC5480 & Sc & & -19.94 & -0.51 & 0.20 & 232 & 1988 L &
Ib & 0.28 & 0.07 & 258 \\
%KPG427B & NGC5679B & Sb & & -21.37 & -0.62 & 0.02 & 1171 & 1982 D
%& II & 0.23 & 0.48 & 345 \\
KPG434B & NGC5746 & SABb & & -21.86 & -1.50 & 0.828 & 183 & 1983 P
& Ia & 0.15 & 0.01 & 259 & LGG386(5) \\
KPG455A & NGC5857 & SBb & & -20.96 & &-0.40 & 5 & 1950 H & & 0.75
& 0.20 & 148 & LGG394(2) \\
KPG455A & NGC5857 & SBb & & -20.96 & &-0.40 & 5 & 1955 M & & 1.03
& 0.30 & 348 \\
KPG536A & NGC6636 & Sc & Sy2 & -20.77 & -0.50 & 0.72 & 427 & 1989
P & Ia & 1.16 & 1.90 & 29 \\
KPG570B & NGC7339 & SABbc & & -19.68 & & -0.10 & 54 & 1989 L & II
n & 0.44 & 0.12 & 317 \\
KPG68B & IC1801 & SBb & & -19.66 & & -0.57 & 189 & 1976 H? & &
0.49 & 0.26 & 317 & LGG61(5) \\
%KPG72AB & M+06-06-62 & SABd & & -21.28 & -0.94 & & 658 & 1961 P &
%Ia & 0.59 & 2.86 & 5 & LGG65  \\
KPG9B & M+02-02-09 & S? & & -19.97 & & & 97 & 1990 ah & II & 0.28
& 0.19 & 59 \\
RRPG70B & NGC1187 &SBc & & -20.24 & -0.37 & & 45 & 1982 R & Ib &
0.54 & 0.25 & 126 \\
RRPG75A & NGC1316 & SAB0 & L & -22.11& -1.50 & 0.77 & 155 & 1980 N
& Ia & 0.71 & 0.57 & 93 & LGG94(2) \\
RRPG75A & NGC1316 & SAB0 & L & -22.11 & -1.50 & 0.77 & 155 & 1981
D & Ia & 0.34 & 0.26 & 351 \\
RRPG85A & NGC1532 & SBb & & -20.89 & -1.26 & 0.52 & 436 & 1981 A&
II & 0.53 & 0.02 & 65 & LGG111(3) \\
RRPG132A & NGC2207 & SABbc & & -21.63 & & -0.23 & 109 & 1975 A &
Ia & 0.98 & 0.95 & 229 \\
RRPG206B & NGC4039 & Sm & & -21.12 & & -0.10 & 9 & 1921 A & & 0.32
& 0.39 & 46 & LGG263(9) \\
RRPG206A & NGC4038 & SBm & & -21.36 & & 0.10 & 9 & 1974 E & & 0.65
& 1.03 & 219 & LGG263(8) \\
RRPG242A & NGC5090 & E2 & & -21.17 & & 0.12 & 216 & 1981 C?& &
0.29 & 0.33 & 173 & LGG339(8) \\
%RRPG243B & E270-G05 & SBab & & -20.79 & & 0.13 & 1292 & 1983 X?& &
%0.79 & 0.10 & 34\\
\hline
\end{tabular}
\end{tiny}
\end{minipage}
\end{table*}

Finally, we have identified SNe discovered in groups of galaxies.
The group definition was taken according to the "hard criterion",
i.e. Huchra-Geller percolation, Tully hierarchical and 2D wavelet
methods (Garcia 1993). The crossing produced a list of 211 SNe
discovered in 170 galaxies members of 116 groups (Table 3). In
Table 3 Columns 1, 2 and 3 report Garcia's designations, group
member counts and mean morphological class (MMC) of groups. The
MMC was calculated as MMC = $\sum X_{i}/N$, where $N$ is the
number of galaxies in the groups and $Xi$ is a code describing the
morphological type of $i$th galaxy. Elliptical galaxies correspond
to 1, S0 to 2, Spirals and Irregulars to 3 and 4, respectively. In
Column 4 is the standard deviation of heliocentric radial
velocities of group members expressed in km $s^{-1}$. Columns 5,
6, 7, 8 and 9 respectively are the same as Columns 2, 3, 4, 5 and
6 of Table 1. The logarithmic ratio of the blue luminosity of the
SN host galaxy to that of the most luminous member of the group is
in Column 10. Column 11 is the ratio of the distance of the SN
host galaxy from the geometrical center of  the group to the group
radius. The latter computed as the distance of the most distant
member from the geometrical center of the group. Columns 12, 13
and 14 are the same with Columns 7, 8 and 9 of Table 1. In Column
15 of Table 3 is the position angle (PA) of the SN with respect to
the line connecting the geometrical center of group to the SN
host. PA is calculated clockwise with the geometrical center
placed at $180^{\circ}$. Column 16 are notes.

In this sample 65 SNe are classified as type Ia, 84 SNe as type
II+Ib/c and 62 SNe are unclassified. 59\% of SNe are in spirals
hosts with barred structure. 14\% of SNe hosts have active nuclei
(18 \% if we include Liners). 36\% of SNe occurred in the most
luminous galaxies of the groups.

\begin{table*}
\centering
\begin{minipage}{170mm}
\begin{tiny}
\caption{Known SN hosts in groups member galaxies}
\begin{tabular}{@{}lllllllccccllccl@{}}
 \hline
  \medskip
  Cat.Name & N & MMC & Sigma & Name  & Type & AGN & $M_{B}$ &
  Log$\frac{L_{FIR}}{L_{B}}$ & Log$\frac{L_{P}}{L_{LG}}$ & $\frac{R_{P}}{R_{GR}}$& SN & & $\frac{R_{0}}{R_{25}}$ & PA &
  Notes\\

 (1) & (2) & (3) & (4) & (5) & (6) & (7) & (8) & (9) & (10) &
  (11) & (12) & (13) & (14) & (15) & (16) \\
  \hline

LGG 2 & 5 & 3.00 & 57 & NGC  23 & SBa & & -21.73 & -0.13  & 0
&0.20& 1955 C? &   & 0.29  &  315 \\
LGG 11 & 3  & 3.00 & 230 & NGC 224 & Sb& & -21.69 & -3.37  & 0&
1.00 & 1885 A & I  & 0.01 & 160\\
LGG 14 & 18 &2.83 & 185 &IC 43 & SABc & & -20.58 & -0.72 & -1.11&
0.66 & 1973 U & II & 0.58 & 178 \\
LGG 14 & 18 & 2.83 &185 & M+05-02-42 & Sba & &-20.36 & & -0.80 &
0.70 & 1954 ac? & &0.78 & 216 \\
LGG 14 & 18 & 2.83 & 185 & M+05-03-16 & S & &-20.67 & -0.43 &
-0.67 & 0.67 & 1990 aa & Ic & 0.63& 83 \\
LGG 18 & 14 & 2.57 & 151 & M+05-03-75 & S & & -20.85 & -0.86 &
-0.48 & 0.15 & 1961 M& & 0.65 & 187 \\
LGG 21 & 5 & 3.40 & 89 & NGC 488 & Sb & & -21.71 & -1.21 & 0 &
0.92 & 1976 G & & 0.73 & 354 \\
LGG 26 & 21 & 2.76 & 182 & NGC 536 & SB & & -22.06 & -1.23 & -0.07
& 0.09 & 1963 N & II & 0.35 & 151 \\
LGG 32 & 3 & 2.67 & 144 & IC1731 & SABc & & -19.90 & -0.24 & -0.59
& 1.00 & 1983 R & Ia & 0.46 & 317 \\
LGG 37 & 18 & 2.44 & 191 & NGC 735 & Sb & & -20.82 & -0.93 & -0.60
& 0.80 & 1972 L & & 0.69 & 81 \\
LGG 37 & 18 & 2.44 & 191 & NGC 753 & SABbc & & -21.74 & -0.36 &
-0.24 & 0.49 & 1954 E & & 0.82 &28 \\
LGG 56 & 7 & 3.43 & 70 & NGC 908 & Sc & & -21.26 & -0.59 & 0 &
0.40 & 1994 ai & Ic & 0.14 & 151 \\
LGG 57 & 3 & 3 & 87 & UGC1867 & Scd & & -21.05 & & 0 & 1.00 & 1989
ac? & & 0.70 & 112 \\
LGG 58 & 4 & 2.75 & 134 & M+05-06-51 & SABab & & -20.29 & & -0.41
& 0.45 & 1965 K & & 0.55 & 18 \\
LGG 59 & 3 & 3.33 & 71 & M-02-07-010 & Sc& & -18.57 & -0.41 &
-0.09 & 0.22 & 1985 S & II & 0.72 & 283 \\
LGG 61 & 7 & 2.71 & 100 & IC1801 & SBb & & -19.67 & & -0.57 &
0.62 & 1976 H?& & 0.49 &306 & KPG68B \\
LGG 61 &7 & 2.71 & 100 & NGC 930 & Sa & & -20.94 & & -0.06 & 0.42&
1992 bf & I & 0.25 & 303 \\
LGG 63 & 6 & 3 & 125 & NGC 977 & SABa & & -19.93 & & -0.58 & 0.52
& 1976 J & Ia & 0.46 & 85 \\
LGG 65 & 3 & 3.00 & 62 & M+06-06-62 & SABd & & -21.28 & -0.94 & 0
& 0.62 & 1961 P & Ia & 0.80 & 155& KPG72 \\
LGG 66 & 12 & 2.83 & 153 & M+05-07-29 & Sc & & -20.42 & -0.54 &
-0.75 & 0.44 & 1982 V & I & 0.27 & 239 \\
LGG 66 & 12 & 2.83 & 153 & M+06-06-68 & SBa & & -20.65 & 0.09 &
-0.66 & 0.61 & 1938 A & I & 0.83 & 164 \\
LGG 70 & 5 & 3.20 & 129 & NGC1003 & Scd & & -19.42 & -1.44 & -0.40
& 0.90 & 1937 D & Ia & 0.30 & 253 \\
LGG  71 & 13 & 3 & 91 & NGC 991 & SABc & & -18.75 & -0.80 & -0.82
& 0.11 & 1984 L & Ib & 0.48 & 254 \\
LGG 71 & 13 & 3 & 91 & NGC1035 & Sc & & -19.22 & -0.54 & -0.63 &
0.13 & 1990 E & II & 0.29 & 337 \\
LGG 71 & 13 & 3 & 91 & NGC1084 & Sc & & -20.56 & -0.20 & -0.09 &
0.13 & 1963 P & Ia & 0.36 & 120 \\
LGG 71 & 13 & 3 & 91 & NGC1084 & Sc & & -20.56 & -0.20 & -0.09 &
0.13 & 1996 an & II & 0.49 & 190 \\
LGG 73 & 5 & 3 & 85 & NGC1073 & SBc & & -19.79 & -1.12 & -0.78 &
0.14 & 1962 L & Ic & 0.53 & 34 \\
LGG 75 & 4 & 2.50 & 84 & NGC1097 & SBb & Sy1&  -21.20 & -0.31 & 0
& 0.22 &1992 bd & II & 0.04 & 357 \\
LGG 86 & 4 & 3 & 48 & NGC1255 & SABbc & & -20.48 & -0.80 & 0 &
0.74 & 1980 O & II&  1.12 & 221 \\
LGG 88 & 11 & 2.09 & 159 & NGC1275 &CD &Sy2 & -23.03&  -0.71 & 0 &
0.40 & 1968 A & I&  0.38 & 111 \\
LGG 94 & 15 & 2.80 & 106 & NGC1310 & SBcd & & -19.02 & -0.73 &
-1.23 & 0.98 & 1965 J & & 0.27 & 358 \\
LGG 94 & 15 & 2.80 & 106 & NGC1316 & SAB0 & L & -22.11 & -1.50 & 0
& 0.89 & 1980 N & Ia & 0.71 & 25 & RRPG75A \\
LGG 94 & 15 & 2.80 & 106 & NGC1316 & SAB0 & L & -22.11 & -1.50 & 0
& 0.89 & 1981 D & Ia & 0.33 & 289 \\
LGG 94 & 15 & 2.80 & 106 & NGC1350 & SBab & & -21.16 & -1.55 &
-0.38 & 0.63 & 1959 A & & 0.52 & 355 \\
LGG 94 & 15 & 2.80 & 106 & NGC1365 & SBb & Sy1.8 & -21.72 & -0.06
& -0.16 & 0.13 & 1957 C & & 0.47 & 231 \\
LGG 94 & 15 & 2.80 & 106 & NGC1365 & SBb & Sy1.8 & -21.72 & -0.06&
-0.16 & 0.13 & 1983 V & Ic & 0.24 & 313 \\
LGG 94 & 15 & 2.80 & 106 & NGC1380 & S0 & & -21.05 & -1.46 & -0.42
& 0.33 & 1992 A& Ia & 0.46 & 315 \\
LGG 96 & 32 & 2.22 & 166 & M-06-09-04 & SBa & & -17.56 & & -1.38 &
0.38 & 1969 A? & & 0.38 \\
LGG 97 & 24 & 2.58 & 146 & NGC1332 & S0 & & -20.24 & -1.59 & -0.38
&  0.77 & 1982 E & & 3.02 & 10 \\
LGG 97 & 24 & 2.58 & 146 & NGC1325 & Sbc & & -20.07 & -1.28 &
-0.45 & 0.89 & 1975 S & II& 0.95 & 44 \\
LGG 102 & 3 & 2.67 & 99 & NGC1411 & S0 & & -18.21 & -1.61 & -0.82
& 0.27 & 1976 L & I & 3.41 & 338 \\
LGG 102 & 3 & 2.67 & 99 & NGC1448 & Scd & & -20.26 & -0.61 & 0 &
1.00 & 1983 S & II & 0.39 & 171 \\
LGG 107 & 3 & 3.33 & 5 & NGC1511 & Sa & & -19.93 & -0.06 & 0 &
0.47 & 1935 C & & 0.84 & 186 \\
LGG 111 & 6 & 2.50 & 113 & NGC1532 & SBb & & -20.89 & -1.04 & 0 &
0.56 & 1981 A & II & 0.53 & 102 & RRPG85A \\
LGG 118 & 6 & 2.83 & 141 & NGC1667 & SABc & Sy2 & -21.75 & -0.25 &
0 & 0.42 & 1986 N & Ia & 0.31 & 351 \\
LGG 127 & 4 & 3.00 & 101 & NGC1808 & SABb & Sy2 & -19.90 & 0.26&
-0.27 & 0.19 & 1993 af & Ia & 1.94 & 302 \\
LGG 145 & 6 & 2.83 & 135 & NGC2268 & SABbc & & -21.20 & -0.70 & 0
& 0.71 & 1982 B & Ia & 0.29 & 38 \\
LGG 152 & 4 & 3.00 & 83 & NGC2487 & SBb & & -21.32 & -1.02 & 0 &
0.30 & 1975 O & Ia & 0.44 & 257 & KPG150B \\
LGG 154 & 5 & 2.60 & 96 & NGC2441 & SABb & & -20.99 & -0.90 &
-0.23 & 0.70 & 1995 E & Ia & 0.38 & 294 \\
LGG 156 & 3 & 3.00 & 120 & M+04-20-13 & SBc & & -20.55 & -0.69 &
-0.16&  0.08 & 1962 F & & 0.89 & 129 \\
LGG 156 & 3 & 3.00 & 120 & NGC2565 & SBbc & & -20.95 & -1.02 & 0 &
0.93 & 1960 M & I & 0.66 & 317 \\
LGG 156 & 3 & 3.00 & 120 & NGC2565 & SBbc & & -20.95 & -1.02 & 0 &
0.93 & 1992 I & II & 1.03 & 12 \\
LGG 165 & 5 & 3.00 & 66 & NGC2715 & SABc & & -21.03 & -1.34 &
-0.07 & 1.00 & 1987 M & Ic & 0.14 & 257 \\
LGG 169 & 3 & 3.00 & 69 & NGC2775 & Sab & & -20.56 & -1.16 & 0 &
0.50 & 1993 Z & Ia & 0.40 & 250 & KIG309 \\
LGG 176 & 5 & 3.20 & 70 & NGC3031 & I0 & & -21.56 & -2.50 & 0 &
0.27 & 1993 J & IIb & 0.21 & 255 & KPG218A \\
LGG 182 & 9 & 3.11 & 44 & M+01-25-25 & SBd & & -19.02 & & -0.38 &
0.75 & 1989 C & II p & 0.07 & 246 \\
LGG 192 & 5 & 3.00 & 91 & NGC3169 & Sa & L & -20.47 & -0.78 & 0 &
0.34 & 1984 E & II & 0.68 & 71 & KPG228A \\
LGG 193 & 4 & 3.25 & 134 & NGC3147 & Sbc & Sy2 & -22.15 & -0.65 &
0 & 0.60 & 1972 H & I & 0.46 & 78 \\
LGG 193 & 4 & 3.25 & 134 & NGC3147 & Sbc & Sy2&  -22.15 & -0.65 &
0 & 0.60 & 1997 bq & Ia & 0.61 & 338 \\
LGG 194 & 10 & 2.60 & 114 & NGC3177 & Sb & & -18.69 & 0.01 & -0.60
& 0.30 & 1947 A & II & 1.04 & 296 \\
LGG 194 & 10 & 2.60 & 114 & NGC3226 & E2 & L & -19.25 & & -0.37&
0.73 & 1976 K & I & 0.37 & 169 & KPG234A \\
LGG 194 & 10 & 2.60 & 114 & NGC3227 & SABc & Sy1.5 & -20.18 &
-0.72 & 0 & 0.75 & 1983 U & Ia & 0.16 & 309 & KPG234B \\
LGG 197 & 5 & 2.40 & 51 & NGC3254 & Sbc & & -20.04 & -1.63 & 0 &
0.86 & 1941 B & & 0.42 & 154 \\
LGG 207 & 5 & 3.00 & 63 & UGC5695 & S & & -18.60 & & -0.59 & 1.00
& 1993 N & II n & 0.65 & 143 \\
LGG 207 & 5 & 3.00 & 63 & UGC5695 & S & & -18.60 & & -0.59 & 1.00
& 1994 N & II&  0.72 & 98 \\
LGG 211 & 14 & 2.50 & 170 & NGC3336 & Sc & & -21.16 & -0.69 &
-0.14 & 0.75 & 1984 S & & 0.17 & 341 \\
LGG 214 & 4 & 3.00 & 39 & NGC3389 & Sc & & -19.70 & & -0.21 & 1.00
& 1967 C & Ia & 0.93 & 263 \\
LGG 216 & 4 & 2.75 & 48 & NGC3367 & SBc & Sy & -21.27 & -0.44 & 0
& 1.00 & 1986 A & Ia & 0.33 & 24 \\
LGG 216 & 4 & 2.75 & 48 & NGC3367 & SBc & Sy & -21.27 & -0.44 & 0
& 1.00 & 1992 C & II & 0.52 & 331 \\
LGG 219 & 5 & 3.20 & 72 & NGC3370 & Sc & & -19.68 & -0.74 & 0 &
1.00 & 1994 ae & Ia & 0.48 & 175 \\
LGG 226 & 3 & 2.67 & 101 & NGC3458 & SAB & & -19.41 & & -0.17 &
1.00 & 1991 F & Ia & 0.58 & 146 \\
LGG 231 & 4 & 3.00 & 99 & NGC3627 & SABb & Sy2 & -21.12 & -0.73&
-0.01 & 0.41 & 1973 R & II & 0.24 & 316 \\
LGG 231 & 4 & 3.00 & 99 & NGC3627 & SABb & Sy2 & -21.12 & -0.73 &
-0.01 & 0.41 & 1989 B & Ia & 0.20 & 270 \\
LGG 231 & 4 & 3.00 & 99 & NGC3627 & SABb & Sy2 & -21.12 & -0.73 &
-0.01 & 0.41 & 1997bs & IIn & 0.29 & 60 \\
LGG 232 & 4 & 2.50 & 57 & NGC3625 & SABb & & -19.47 & -1.30 &
-0.57 & 0.17  & 1983 W & Ia & 0.40 & 21 \\
LGG 241 & 8 & 3.00 & 154&  NGC3631 & Sc & & -20.75 & -0.88 & 0 &
1.00 & 1964 A & V & 0.78 & 156 \\
LGG 241 & 8 & 3.00 & 154 & NGC3631 & Sc & & -20.75 & -0.88 & 0 &
1.00 & 1965 L & II & 0.46 & 138 \\
LGG 241 & 8 & 3.00 & 154 & NGC3631 & Sc & & -20.75 & -0.88 & 0 &
1.00 & 1996 bu & II n & 0.80 & 120 \\
LGG 241 & 8 & 3.00 & 154 & NGC3913 & Sd & & -18.04 & -1.19 & -1.08
& 0.49 & 1963 J & Ia & 0.17 & 42 \\
LGG 241 & 8 & 3.00 & 154 & NGC3913 & Sd & & -18.04 & -1.19 & -1.08
& 0.49 & 1979 B & Ia & 0.57 & 227 \\
LGG 247 & 3 & 3.00 & 8 & NGC3780 & Sc & & -20.95 & -0.87 & 0 &
1.00 & 1978 H & II&  0.23 & 298 \\
LGG 247 & 3 & 3.00 & 8 & NGC3780 & Sc & & -20.95 & -0.87 & 0 &
1.00 & 1992 bt & II & 0.24 & 159 \\
LGG 250 & 7 & 3.00 & 125 & NGC3733 & SABcd & & -19.32 & -1.69 &
-0.40 & 0.73 & 1980 D & II & 0.84 & 118 \\
LGG 250 & 7 & 3.00 & 125 & NGC3756 & SABbc & & -20.16 & -1.22&
-0.07 & 0.66 & 1975 T & II & 0.92 & 243 \\
LGG 255 & 5 & 2.00 & 96 & NGC3904 & E2-3 & & -19.87 & & -0.47 &
0.37 & 1971 C  & Ia & 2.18 & 239 \\
LGG 258 & 27 & 2.89 & 118 & NGC3992 & SBbc & & -21.12 & -1.63 & 0
& 0.64 & 1956 A & Ia & 0.35 & 246 \\
LGG 258 & 27 & 2.89 & 118 & NGC4088 & SABbc & & -20.35 & -0.70&
-0.31 & 0.50 & 1991 G & II & 0.50 & 31 \\
LGG 258 & 27 & 2.89 & 118 & NGC4157 & SABb & & -19.83 & -0.62&
-0.51 & 0.70 & 1937 A & II & 0.59 & 238 \\
LGG 258 & 27 & 2.89 & 118 & NGC4157 & SABb & & -19.83 & -0.62 &
-0.51 & 0.70 & 1955 A & & 0.56 & 214 \\
LGG 258 & 27 & 2.89 & 118 & NGC4220 & S0 & & -19.01 & -0.99 &
-0.84 & 0.87 & 1983 O & & 0.24 & 125 \\
LGG 258 & 27 & 2.89 & 118 & UGC6983 & SBcd & & -18.60 & -1.40 &
-1.01 &  0.56 & 1994 P & II & 1.07 & 262 \\
LGG 258 & 27 & 2.89 & 118 & UGC6983 & SBcd & & -18.60 & -1.40 &
-1.01 & 0.56 & 1964 E & Ia & 0.92 & 93 \\
LGG 259 & 6 & 3.00 & 82 & IC2973 & SBd & &  -19.43 & & -0.89 &
1.00 & 1991 A & Ic&  0.77 & 276 \\
LGG 259 & 6 & 3.00 & 82 & NGC3995 & Sm & & -21.66 & & 0 & 0.45 &
1988 ac & & 0.73 & 43 \\
LGG 263 & 12 & 2.58 & 72 & NGC4038 & SBm & & -21.36 & & 0 & 0.39 &
1974 E & & 0.65 & 143 & RRPG206A \\
LGG 263 & 12 & 2.58 & 72&  NGC4039 & Sm & & -21.12 & & -0.10 &
0.39 & 1921 A & & 0.32 & 148 & RRPG206B \\
LGG 266 & 3 & 2.67 & 85 & NGC4041 & Sbc & & -19.94 & -0.31 & -0.20
& 0.23 & 1994 W & II n & 0.24&  296 \\
LGG 269 & 16 &  2.75 & 118 & NGC3938 & Sc & & -19.95 & -0.93 & 0 &
0.77 & 1961 U & II L&  0.81 & 352 \\
LGG 269 & 16 & 2.75 & 118 & NGC3938 & Sc & & -19.95 & -0.93 & 0 &
0.77 & 1964 L & Ic & 0.19 & 36 \\
LGG 269 & 16 & 2.75 & 118 & NGC4051 & SABbc & Sy1.5 & -19.84 &
-0.94 & -0.04 & 0.72 & 1983 I & Ic & 0.38 & 215 \\
LGG 269 & 16 & 2.75 & 118 & NGC4096 & SABc & & -19.84 & -1.26 &
-0.04 & 0.75 & 1960 H & Ia & 0.78 & 146\\
LGG 269 & 16 & 2.75 & 118 & NGC4183 & Scd & & -19.72 & & -0.09 &
0.64 & 1968 U?& &  0.63 &1 \\
LGG 276 & 8 & 2.75 & 73 & NGC4185 & Sbc & & -20.63 & -0.85 & -0.05
& 0.33 & 1982 C & Ia & 0.72 & 36 \\
LGG 279 & 11 & 2.73 & 186 & NGC4136 & SABc & & -18.41 & -1.25 &
-0.69 & 1.00 & 1941 C & II & 0.67 & 305 \\
\end{tabular}
\end{tiny}
\end{minipage}
\end{table*}

\begin{table*}
\centering
\begin{minipage}{170mm}
\contcaption{}
\begin{tiny}
\begin{tabular}{@{}lllllllccccllccl@{}}
  \hline
  (1) & (2) & (3) & (4) & (5) & (6) & (7) & (8) & (9) & (10) &
  (11) & (12) & (13) & (14) & (15) & (16) \\
  \hline
LGG 279 & 11 & 2.73 & 186 & NGC4414 & Sc & & -19.98&  -0.50 &
-0.06 & 0.89 & 1974 G & Ia & 0.57 & 148 \\
LGG 281 & 18 & 2.67 & 132 & NGC4273 & SBc & & -20.78 & -0.24 &
-0.03 & 0.04 & 1936 A & II & 0.42 & 13 \\
LGG 285 & 25 & 2.68 & 134 & NGC4189 & SABcd & & -20.33 & -0.60 &
-1.03 & 0.58 & 1966 E & II & 0.77 & 109 \\
LGG 285 & 25 & 2.68 & 134 & NGC4254 & Sc & & -22.52 & -0.44 &
-0.16 & 0.41 & 1967 H & & 0.51 & 295 \\
LGG 285 & 25 & 2.68 & 134 & NGC4254 & Sc & & -22.52 & -0.44 &
-0.16 & 0.41 & 1972 Q & II P & 0.60 & 30 \\
LGG 285 & 25 & 2.68 & 134 & NGC4254 & Sc & & -22.52 & -0.44 &
-0.16 & 0.41 & 1986 I & II P & 0.24 & 285 \\
LGG 285 & 25 & 2.68 & 134 & NGC4568 & Sc & & -21.61& & -0.52 &
0.76 & 1990 B & Ib/c & 0.16 & 270 & KPG347B \\
LGG 287 & 16 &  2.69 & 183 & NGC4303 & SABbc & Sy & -21.75 & -0.55
& 0 & 0.50 & 1926 A & II & 0.36 & 67 \\
LGG 287 & 16 & 2.69 & 183 & NGC4303 & SABbc & Sy & -21.75 & -0.55
& 0 & 0.50 & 1961 I & II & 0.42 & 320 \\
LGG 287 & 16 & 2.69 & 183 & NGC4303 & SABbc & Sy & -21.75 & -0.55
& 0 & 0.50 & 1964 F & II & 0.14 & 150 \\
LGG 287 & 16 & 2.69 & 183&  NGC4496A & SBm & & -20.20 & & -0.62&
0.24 & 1960 F & Ia & 0.38 & 246 & KPG343A \\
LGG 287 & 16 & 2.69 & 183 & NGC4527 & SABbc & L & -21.35 & -0.31 &
-0.16 & 0.43 & 1915 A & & 0.43 & 144 \\
LGG 287 & 16 & 2.69 & 183 & NGC4527 & SABbc & L & -21.35 & -0.31 &
-0.16 & 0.43 & 1991 T & Ia & 0.54 & 214 \\
LGG 287 & 16 & 2.69 & 183 & NGC4536 & SABbc & & -21.65 & -0.41 &
-0.04 & 0.51 & 1981 B & Ia & 0.60 & 189 \\
LGG 289 & 63 & 2.68 & 217 &  NGC4302 & Sc & & -20.02 & & -0.81 &
0.17 & 1986 E & II & 0.82 & 147 & KPG332B \\
LGG 289 & 63 & 2.68 & 217 & NGC4321 & SABbc & L & -22.05 & -0.75 &
0 & 0.08 & 1901 B & I & 0.56 & 192 \\
LGG 289 & 63&  2.68 & 217 & NGC4321 & SABbc & L & -22.05 & -0.75 &
0 & 0.08 & 1914 A & & 0.55 & 296 \\
LGG 289 & 63 & 2.68 & 217 & NGC4321 & SABbc & L & -22.05 & -0.75 &
0 & 0.08 & 1959 E & I & 0.32 & 354 \\
LGG 289 & 63 & 2.68 & 217 & NGC4321 & SABbc & L & -22.05 & -0.75 &
0 & 0.08 & 1979 C & II & 0.53 & 317 \\
LGG 289 & 63 & 2.68 & 217 & NGC4411B & SABcd & & -18.50 & -1.09 &
-1.42 & 0.66&  1992 ad & II & 0.58 & 46 & KPG336B \\
LGG 289 & 63 & 2.68 & 217 & NGC4486 & E & Sy & -21.96 & -2.61 &
-0.04 & 0.34 & 1919 A & I & 0.40 & 201 \\
LGG 289 & 63  & 2.68 & 217 & NGC4564 & E6 & & -19.29 & & -1.10 &
0.46 & 1961 H & Ia & 0.05 & 204 \\
LGG 289 & 63 & 2.68 & 217 & NGC4579 & SABb & Sy1.9 & -21.49 &
-1.10 & -1.10 & 0.45 & 1988 A & II P & 0.33 & 29 \\
LGG 289 & 63 & 2.68 & 217 & NGC4579 & SABb & Sy1.9 & -21.49 &
-1.10 & -1.10 & 0.45 & 1989 M & Ia & 0.28 & 272 \\
LGG 289 & 63 & 2.68 & 217 & NGC4639 & SABbc & Sy1.8 & -19.07 &
-1.05 & -1.19 & 0.44 & 1990 N & Ia & 0.87 & 143 \\
LGG 289 & 63 & 2.68 & 217 & NGC4647 & SABc & & -19.93 & -0.55 &
-0.85 & 0.54 & 1979 A & I & 0.68 & 280 & KPG353A \\
LGG 290 & 17 & 3.41 & 142 & NGC4258 & SABbc & Sy1.9 & -21.13 & &0
& 0.85 & 1981 K & II & 0.26 & 150 \\
LGG 290 & 17 & 3.41 & 142 & NGC4490 & SBd & & -20.67 & -0.83 &
-0.19 & 0.23 & 1982 F & II P & 0.22 & 237 & KPG341B \\
LGG 290 & 17 & 3.41 & 142 & NGC4618 & SBm & & -19.04 & -1.17 &
-0.84 & 0.31 & 1985 F & Ib & 0.13 & 242 & KPG349A \\
LGG 291 & 6 & 3.33 & 169 & NGC4214 & IABm & & -17.18 & &-1.90 &
0.85 & 1954 A & Ib & 0.92 & 257 \\
LGG 292 & 59 & 2.42 & 152 & NGC4340 & SB0 & & -18.74 & & -1.09 &
0.22 & 1977 A & & 0.51 & 110 \\
LGG 292 & 59 & 2.42 & 152 & NGC4374 & E1 & & -20.72 & -2.32 &
-0.30 & 0.46 & 1957 B & Ia & 0.25 & 161 \\
LGG 292 & 59 & 2.42 & 152 & NGC4374 & E1 & & -20.72 & -2.32 &
-0.30 & 0.46 & 1991 bg & Ia&  0.30 & 334 \\
LGG 292 & 59 & 2.42 & 152 & NGC4374 & E1 & & -20.72 & -2.32 &
-0.30 & 0.46 & 1980 I & Ia & 2.35 & 64 \\
LGG 292 & 59 & 2.42 & 152 & NGC4382 & S0 & & -20.43 & & -0.42 &
0.21 & 1960 R & Ia & 0.63 & 238 & KPG334A \\
LGG 292 & 59 & 2.42 & 152 & NGC4636 & E/S0 & L & -20.74 & & -0.29
& 0.46 & 1939 A & Ia & 0.18 & 20 \\
LGG 292 & 59 & 2.42 & 152 & NGC4688 & SBcd & & -17.79 & -0.82 &
-1.47 & 0.66 & 1966 B & II & 0.47 & 115 \\
LGG 292 & 59 & 2.42 & 152 & NGC4772 & Sa & L & -19.23 & & -0.89 &
0.62 & 1988 E & & 0.270 & 303 \\
LGG 295 & 4 & 2.50 & 130 & NGC4512 & SBdm & & -18.99 & -0.99 &
-0.65 & 0.16 & 1995 J & II & 0.79 & 51 & KPG345A \\
LGG 295 & 4 & 2.50 & 130 & NGC4545 & SBcd & & -20.61 & -0.99 & 0 &
1.00 & 1940 D & & 0.36 & 246 \\
LGG 298 & 54 & 2.48 & 248 & NGC4650A & S & & -19.51&  -0.84 &
-0.94 & 0.07 & 1990 I & I b & 1.42 & 136 \\
LGG 299 & 3 & 3.00 & 100 & NGC4632 & Sc & & -20.59 & -0.79 & -0.05
& 1.00 & 1946 B & II & 0.19 & 23 \\
LGG 299 & 3 & 3.00 & 100 & NGC4666 & SABc & L & -20.71 & -0.09 & 0
& 0.40 & 1965 H & II P & 0.28 & 45 \\
LGG 307 & 14 & 3.07 & 188 & NGC4699 & SABb & Sy & -21.22 & -1.02 &
0 & 0.95 & 1948 A & & 0.50 & 175 \\
LGG 307 & 14 & 3.07 & 188 & NGC4699 & SABb & Sy & -21.22 & -1.02 &
0 & 0.95 & 1983 K & II P & 1.75 & 126 \\
LGG 314 & 18 & 3.11 & 160 & NGC4948 & SBdm & & -18.37 & & -1.01 &
0.86 & 1994 U & Ia & 0.780 \\
LGG 315 & 12 & 3.08 & 215 & NGC4753 & I0 & & -20.50 & -1.18 & 0 &
0.32 & 1965 I & Ia & 0.66 & 232 \\
LGG 315 & 12 & 3.08 & 215 & NGC4753 & I0 & & -20.50 & -1.18 & 0 &
0.32 & 1983 G & Ia & 0.12 & 306 \\
LGG 316 & 3 & 1.67 & 137 & NGC4783 & E & & -21.86 & & 0 & 1.00 &
1956 B & & 0.78 & 152 \\
LGG 321 & 3 & 3.00 & 48 & NGC4902 & SBb & & -21.26 & -0.73 & 0 &
1.00 & 1979 E? & & 0.76 & 189 \\
LGG 321 & 3 & 3.00 & 48 & NGC4902 & SBb & & -21.26 & -0.73 & 0 &
1.00 & 1991 X & Ia & 0.20 & 135 \\
LGG 333 & 5 & 3.00 & 112 & NGC4981 & SABbc & L & -19.82&  -0.52 &
-0.24 & 0.57 & 1968 I & Ia & 0.09 & 312 \\
LGG 334 & 9 & 3.22 & 110 & NGC5033 & Sc & Sy1.9 & -20.96 & -0.94 &
0 & 0.36 & 1950 C & & 1.44 & 229 \\
LGG 334 & 9 & 3.22 & 110 & NGC5033 & Sc & Sy1.9 & -20.96 & -0.94 &
0 & 0.36 & 1985 L & II  &0.44 & 23 \\
LGG 335 & 3 & 3.00 & 26 & NGC5020 & SABbc & & -21.15 & -0.39 & 0 &
1.00 & 1991 J & II & 0.77 & 119 \\
LGG 339 & 17 & 2.29 & 199 & E323-G99 & SABc & & -20.27 & -0.49 &
-0.50 & 0.34 & 1984 I & Ib & 0.72 & 0 \\
LGG 339 & 17 & 2.29 & 199 & NGC5090 & E2 & & -21.17 & & -0.14 &
0.83 & 1981 C? & & 0.29 & 104 & RRPG242A \\
LGG 342 & 8 & 2.75 & 125 & E269-G57&  SABb & & -21.40 & & -0.01 &
0.96 & 1992 K & Ia P & 0.19 & 242 \\
LGG 344 & 7 & 3.14 & 108 & NGC5128 & S0 & Sy2 & & & & 0.41 & 1986
G & Ia & 0.22 & 35 \\
LGG 347 & 4 & 3.25 & 135 & NGC5055 & Sbc & L & -21.15 & -1.14 & 0
& 0.75 & 1971 I & Ia & 0.68 & 182 \\
LGG 347 & 4 & 3.25 & 135 & NGC5194 & Sbc & Sy2 & -20.50 & -0.87 &
-0.26 & 0.87 & 1994 I & Ic & 0.07 & 74 & KPG379A \\
LGG 348 & 3 & 2.00 & 74 & NGC5082 & SB0 & & -20.43 & & 0 & 0.79 &
1958 F? & & 0.80 & 173 \\
LGG 350 & 5 & 2.60 & 182 & NGC5127 & E & & -21.08 & & 0 & 0.24 &
1991 bi & Ia & 0.17 & 193 \\
LGG 355 & 3 & 3.67 & 106 & NGC5236 & SABc & & -22.25 & -0.88 &
-0.13 & 1.00 & 1923 A & II & 0.32 & 55 \\
LGG 355 & 3 & 3.67 & 106 & NGC5236 & SABc & & -22.25 & -0.88 &
-0.13 & 1.00 & 1945 B & & 0.52 & 268 \\
LGG 355 & 3 & 3.67 & 106 & NGC5236 & SABc & & -22.25 & -0.88 &
-0.13 & 1.00 & 1950 B & & 0.27 & 207 \\
LGG 355 & 3 & 3.67 & 106 & NGC5236 & SABc & & -22.25 & -0.88 &
-0.13 & 1.00 & 1957 D & & 0.33 & 108 \\
LGG 355 & 3 & 3.67 & 106 & NGC5236 & SABc & & -22.25 & -0.88 &
-0.13 & 1.00 & 1968 L & II & 0.01&  207 \\
LGG 355 & 3 & 3.67 & 106 & NGC5236 & SABc & & -22.25 & -0.88 &
-0.13 & 1.00 & 1983 N & Ib & 0.42 & 254 \\
LGG 361 & 13 & 2.92 & 122 & NGC5371 & SABbc & L & -21.88 & -0.91 &
0 & 0.62 & 1994 Y & II n & 0.29 & 350 \\
LGG 364 & 3 & 2.67 & 127 & NGC5378 & SBa & & -19.51 & & -0.29 &
0.46 & 1991 ak & Ia & 0.50 & 30 \\
LGG 371 & 3 & 3.00 & 40 & NGC5457 & SABcd & & -20.87 & -0.49 & 0 &
0.51 & 1909 A & II & 0.87 & 167 & KIG610 \\
LGG 371 & 3 & 3.00 & 40 & NGC5457 & SABcd & & -20.87 & -0.49 & 0 &
0.51 & 1951 H & II & 0.41 & 27 \\
LGG 371 & 3 & 3.00 & 40 & NGC5457 & SABcd & & -20.87 & -0.49 & 0 &
0.51 & 1970 G & II L & 0.45 & 151 \\
LGG 372 & 6 & 2.67 & 82 & NGC5480 & Sc & & -19.94 & -0.52 & -0.34
& 0.49 & 1988 L & Ib & 0.28 & 343 & KPG416A \\
LGG 373 & 6 & 2.50 & 123 & NGC5485 & S0 & & -19.66 & & -0.20 &
0.36 & 1982 W & Ia & 0.93 & 273 \\
LGG 374 & 4 & 2.75 & 108 & NGC5493 & S0 & & -20.68 & & -0.20 &
1.00 & 1990 M & Ia & 0.33 & 9 \\
LGG 374 & 4 & 2.75 & 108 & NGC5426 & Sc & & -20.57 & & -0.24 &
0.57 & 1991 B & Ia & 0.27 & 273 \\
LGG 374 & 4 & 2.75 & 108 & NGC5427 & Sc & Sy2 & -21.17 & -0.49 & 0
& 0.55 & 1976 D & Ia & 0.65 & 71 \\
LGG 381 & 3 & 3.00 & 44 & NGC5548 & S0/a & Sy1.5 & -21.55 & -0.90
& 0 & 0.71 & 1984 Z? & II&  0.03\\
LGG 382 & 3 & 2.67 & 66 & M+02-37-15a & SBb & & -19.97 & -0.68 &
-0.30 & 1.00 & 1988 F & Ia & 0.97 & 238 \\
LGG 386 & 22 & 2.95 & 114 & NGC5668 & Sd & & -19.88 & -0.81 &
-0.79 & 0.58 & 1952 G & & 0.32 & 175 \\
LGG 386 & 22 & 2.95 & 114 & NGC5668 & Sd & & -19.88 & -0.81 &
-0.79 & 0.58 & 1954 B & Ia & 0.20 & 207 \\
LGG 386 & 22 & 2.95 & 114 & NGC5746 & SABb & & -21.86 & -1.53 & 0
& 0.41 & 1983 P & Ia & 0.15 & 131 & KPG434B \\
LGG 390 & 3 & 2.33 & 67 & M-03-38-25 & SABbc & & -19.96 & -0.79 &
-0.69 & 1.00 & 1994 V & & 0.49 & 273 \\
LGG 393 & 8 & 1.88 & 147 & NGC5854 & SB0 & & -19.52 & & -0.69 &
0.49 & 1980 P & I & 0.16 \\
LGG 394 & 3 & 3.00 & 134 & NGC5857 & SBb & & -20.96 & & -0.40 &
0.48 & 1950 H & & 0.76 & 107 & KPG455A \\
LGG 394 & 3 & 3.00 & 134 & NGC5857 & SBb & & -20.96 & & -0.40 &
0.48 & 1955 M & & 1.03 & 308 \\
LGG 395 & 4 & 3.00 & 111 & NGC5905 & SBb & Sy1 & -21.05 & -0.58 &
-0.14 & 0.76 & 1963 O & I & 0.61 & 40 \\
LGG 396 & 3 & 2.67 & 59 & NGC5879 & Sbc & L & -19.40 & -1.10 &
-0.68 & 0.53 & 1954 C & II & 0.36 & 139 \\
LGG 396 & 3 & 2.67 & 59 & NGC5907 & Sc & & -21.11 & -1.37 & 0 &
1.00 & 1940 A & II L & 0.92 & 112 \\
LGG 422 & 22 & 2.36 & 182 & IC4798 & S0 & & -21.36 & & -0.40 &
0.60 & 1971 R & & 0.24 & 71 \\
LGG 427 & 12 & 2.50 & 184 & IC4831 & SABb & & -22.15 & -1.40 & 0 &
0.33 & 1992 N & II & 1.02 & 44 \\
LGG 429 & 8 & 2.63 & 115 & IC4963 & Sba & & -20.27 & 0.00 & -0.79
& 0.82 & 1987 I & Ia & 0.14 & 256 \\
LGG 436 & 3 & 3.00 & 48 & NGC6907 & SBbc & & -21.91 & -0.31 & 0 &
1.00 & 1984 V? & & 0.56 & 110 \\
LGG 439 & 4 & 3.00 & 87 & UGC11616 & Scd & & & & & 0.96 & 1991 au
& II & 0.47 & 98 \\
LGG 441 & 11 & 2.27 & 133 & NGC7038 & Sc & & -22.32 & -0.88 & 0 &
0.84 & 1983 L & Ia & 0.38 & 151 \\
LGG 459 & 4 & 3.00 & 24 & NGC7331 & Sb & L & -21.40 & -0.93 & 0 &
0.45 & 1959 D & II L & 0.27 & 84 \\
LGG 465 & 3 & 2.67 & 60 & NGC7418A & Sd & & -18.31 & -0.64 & -0.97
& 0.74 & 1983 M? & & 0.96 &98 \\
LGG 466 & 5 & 2.67 & 80 & IC5270 & SBc & & -19.99 & -0.70 & -0.30
& 0.35 & 1993 L & Ia & 0.53 & 69 \\
LGG 469 & 5 & 2.00 & 99 & NGC7448 & Sbc & & -21.18 & -0.58 & 0 &
0.75 & 1980 L & & 0.34 & 206 \\
LGG 469 & 5 & 2.00 & 99 & NGC7448 & Sbc & & -21.18 & -0.58 & 0 &
0.75 & 1997 dt & Ia & 0.18 & 209 \\
LGG 473 & 15 & 2.20 & 170 & NGC7619 & E & & -21.82 & & 0 & 0.16 &
1970 J & Ia & 0.34 & 132 \\
LGG 473 & 15 & 2.20 & 170 & NGC7634 & SB0 & & -19.97 & & -0.74 &
0.21 & 1972 J & Ia & 1.12 & 157 \\
LGG 476 & 4 & 2.75 & 37 & IC1501 & SABbc & &-20.35 & -0.61 & 0 &
0.51 & 1993 ad & II & 0.65 & 98 \\
LGG 478 & 3 & 3.33 & 31 & NGC7713 & SBd & & -18.41 & & -0.05 &
0.48 & 1982 L & II & 0.65 & 84 \\
LGG 480 & 5 & 3.00 & 98 & NGC7723 & SBb & & -20.55 & -0.70 &
-0.067 & 0.59 & 1975 N & Ia & 0.56 & 200

\end{tabular}
\end{tiny}
\end{minipage}
\end{table*}

It is worth noting again that, because of different selection
criteria, the KIG and KPG (+RRPG) samples have significant
intersection with the Garcia's groups. In particular, about 50\%
of the isolated pairs are the seeds of larger LGG aggregations.

\begin{table*}
\centering
\caption{Mean parameters (and standard deviation) of the SNe host galaxies}
\begin{tabular}{ccccccc}
\hline
       &  \multicolumn{2}{c}{ KIG hosts} & \multicolumn{2}{c}{ KPG+RRPG hosts}
       & \multicolumn{2}{c}{ LGG hosts}\\
\hline
         &    Ia           &   II+Ib/c       &     Ia          & II+Ib/c       
 & Ia              &  II+Ib/c\\
 $<M_B>$ & $-20.56\pm0.20$ & $-20.36\pm0.86$ & $-21.41\pm0.74$ & $-20.57\pm0.98$ 
 & $-20.50\pm0.98$ & $-20.44\pm1.13$ \\
 $<T> $  & $ 1.9\pm3.3 $ & $4.7\pm2.0 $  & $1.7\pm2.6$          & $5.0\pm1.4$ 
 &  $1.9\pm3.5 $ & $4.1\pm 1.4$\\
\hline
\end{tabular}
\end{table*}

\section{ Analysis and discussion}
\subsection{The properties of the host galaxies}

Table 4 presents the mean values and standard deviations of the
absolute blue magnitudes ($<M_B>$) and morphological types ($<T>$)
of SNe host galaxies for the different systems.

It appears that there are not significant differences
(significance level of 95\%) among the means of  blue luminosity
of host galaxies of core collapse SNe in different environments.
Instead, the average absolute blue luminosity  of  parent galaxies
of type Ia SNe in pairs is significantly higher (significant level
p=0.01 or 99.1\%) than in other systems. We could not find an obviuos 
explanation for this finding, though we should be wary of the small statisics.
Indeed of the 9 SN~Ia in paired galaxies, two were discovered in the giant 
galaxy NGC1316 ($M_B=22.11$).

Not significant differences appear in the mean morphological types
of parent galaxies of all types of SNe in different environments.
As expected the morphological types of the hosts of core collapse
SNe is more advanced than those of type Ia. Indeed, type II+Ib/c
SNe occurs only in late type galaxies while SNIa occurs in all
kind of galaxies (Barbon et al. 1999).

\subsection{The location of the SNe inside the host systems}

The small statistics in the samples of KIG and KPG+RRPG does not
allow the comparison of the radial distributions between the two
SNe types in pairs. Where we have sufficient statistics as in the
case of the LGG, a two-sample Kolmogorov-Smirnov (K-S) test does
not show any difference between SNe Ia and SNe II+Ib/c
(significance level of 95\%) in the distributions of the relative
radial distances - $R_0/R_{25}$, corrected for the inclination
along the line of sight (McCarthy 1973). A similar result was
found by Bartunov et al. (1992) who showed that SNIa and SNII in a
general sample of galaxies have similar gradients in the relative
radial distributions and similar scale lengths.

\begin{figure}
   \centering
   \includegraphics[width=9cm]{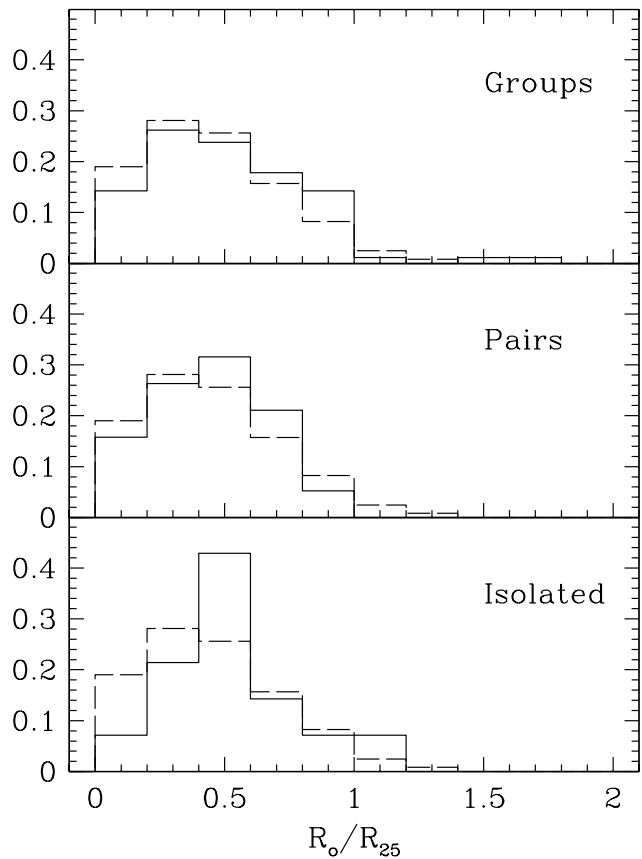}
   \caption{Histograms of the relative radial distributions of the
SNe in isolated parent galaxies (KIG), pairs of galaxies
(KPG+RRPG) and Garcia's groups. The relative radial distances have
been corrected for the inclination along the line of view as
McCarthy (1973).  Long dashed line is the radial distribution in
the reference sample of SNe from Bartunov et al. (1992).}
    \label{histo}
    \end{figure}

We then compared the radial distributions of the core collapse SNe
in the three samples defined in Sect. 2 with that of the reference
sample of Bartunov et al. (1992) containing 121 objects (Figure
\ref{histo}). The two sample K-S test shows that all the distribution come
from the same population indicating that the star forming regions
in galaxies with different environments have the  same radial
distribution.

This result is at variance with the results by Petrosian \&
Turatto (1995) for the core collapse SNe in strongly interacting
systems. A new computation with better statistics has shown that
the radial distribution of the SNeII in  strongly interacting
systems resembles that of our KPG sample. We conclude that if the
interaction triggers star formation in the central regions of the
galaxies, this is too close to the nucleus to be detected by most
current SN searches.

The SNe azimuth distributions in the galaxy systems has been
studied by measuring the SNe position angles relative to the
companions. In galaxy pairs 23 SNe (45.1\%) were located on the
side of the host galaxy pointing to the companion (60\% of SNe
type Ia and 40\% of type II+Ib/c). According to K-S tests the PA
distributions of both types of SNe are not significantly different
from the homogeneous one. Similar result was found in the closely
interacting systems by Petrosian \& Turatto (1995).

A similar test has been performed for the location of SNe in the
host galaxies relative to  the geometrical center of the groups.
54.7\% of type Ia SNe and 50.6\% of type II+Ib/c SNe occur on the
side of the geometrical center of the groups. Again a K-S test
show no preferential direction in the SN distribution. This means
that star formation inside the parent galaxies of SNe is not
affected by local mass concentration. This is not surprising since
we found no influence in the case of interacting neighbors.

The radial distributions  ($R_0/R_{25 }$) of the SNe located on
the side of the companion in galaxy pairs and on the side of the
geometrical center in the groups were compared  with the
distributions on the opposite sides. Again no differences were
found at 95\% significance level.

We investigated also the locations and properties of the host
galaxies of SNe inside the groups. The radial distribution of SN
parent galaxies with respect to the geometrical centers of the
groups  ($R_P/R_{GR}$) does not differ from the radial
distribution of all other members of the groups. The distributions
of the number of galaxies in the groups and the mean morphological
class of the groups hosting SNe was compared with the
distributions of all Garcia (1993) groups.  Once more the
distributions were similar and independent on SN type. We have
therefore evidence that the mean properties of the groups hosting
SNe are not different from others.

The fact that groups with multiple SNe are richer and have in
average later morphological types, is consistent with  the well
known dependence of the SN production on galaxy type and
luminosity.

\subsection{The frequency of SNe in isolated galaxies, in isolated pairs and groups of galaxies}

According to the Tables 1, 2 and 3 the ratios of  SN types,
$n_{Ia}/n_{(II+Ib/c)}$, in isolated parents, in members of pairs
and groups of galaxies are equal to 0.29,  0.54, 0.78
respectively. Taken to face value this suggests a decrease of the
massive progenitors population moving to denser environments. It
is well known however that not all galaxies in the sky are equally
surveyed for the detection of SNe and that selection effects play
a crucial role in the SN discovery (e.g. Cappellaro et al. 1993a).
Therefore, the numbers above should not be taken as representative
of the intrinsic rate of explosions of the different SN types.

We have used the combined archives of five SN searches, namely the
Asiago, Crimea, Calan-Tololo and OCA photographic surveys and the
visual search by Evans, and computed SNe rates in the three
samples following the prescriptions by Cappellaro et al. (1999).
This  computation makes use of the control time method which has
been introduced by Zwicky (1942) and revisited by several authors
(Cappellaro \& Turatto 1988, van den Bergh, McClure \& Evans 1987,
van den Bergh \& McClure 1994, Cappellaro et al. 1993, 1997,
1999).

254 isolated galaxies from KIG, 558 galaxies in pairs from KPG and
RRPG, and 2596 galaxies from Garcia (1993)  groups  were
surveilled in at least one of the afore  mentioned SN searches.
During the surveillance only two SNe were found in the sample of
isolated galaxies, the type Ia SN 1993C and the type II SN 1969B.
17 were found in pairs, while 73 are the SNe detected by the
searches in groups of galaxies.

The rates of SNe for each of the three samples are reported in
Table 5. The rates per unit of blue luminosity are expressed in
SNu, where $1 SNu = 1SN (100yr)^{-1}(10^{10}L_{\sun}^B)$. The
numbers of SNe in each cell of Table 5 are reported in brackets.
Following Cappellaro et al. (1997), unclassified SNe were distributed 
among the three basic types according to the observed distribution 
in the given sample.
In the last column 5 are the average rates after Cappellaro et al.
(1999). Because the same method has been used, these numbers are
directly comparable. The indicated standard errors take into account the
event statistics, the uncertainty on the input parameters and the
bias corrections. The rates for the isolated galaxies are only
nominal values (and no error is reported) since they are based on
a single object per each SN type.

It results that the rate of SNe in pairs is systematically
higher than in groups (by $\sim 60$\%) and in the average sample 
(by $\sim40$\%) though, because of the relatively large uncertainties, 
this is not a solid conclusion. 
This is consistent with the similar finding of Smirnov and Tsvetkov (1981) 
which however did not account for the differences in surveillance time of
the different galaxy samples.

As mentioned above, the fact that SNe are more frequent in the
brighter components of the pairs and groups is consistent with the
fact that in all galaxy types the SNe rate are proportional to the
galaxy luminosity (e.g. Cappellaro et al. 1993a).

\begin{table*}
\centering
\caption{Frequency per luminosity unit $(SNu^\ast)$  with standard errors}
\begin{tabular}{@{}lcccc}
\hline
      SN type & Isolated galaxies  & Galaxies in pairs  & Galaxies in groups  & Cappellaro et al.
      (1999)\\
      \hline
    Ia & 0.14 (1)  &  $0.29\pm0.11$ (8.7)& $0.22\pm0.06$ (42.2)&
    $0.20\pm0.06$(69.6)\\
 II+Ib/c  &  0.28 (1)   &$ 0.66\pm0.31 $(8.3)& $0.37\pm0.17$ (30.8)  & $
 0.48\pm0.19$  (67.4)\\
  ALL  & 0.42 (2) &  $0.95\pm0.32$ (15) &
  $0.59\pm0.18 (73)$ & $0.68\pm0.20$(137)\\
\hline
 \multicolumn{4}{c}{$\ast 1 SNu = 1SN (100yr)^{-1}(10^{10}L_{\sun}^B); H_0=75 km s^{-1}Mpc^{-1}$}
\end{tabular}
\end{table*}

\subsection{Multivariate Factor Analysis}

As a further, general exploration we applied the Multivariate
Factor Analysis (MFA) method to our samples. The MFA is a
statistical method for detection of correlation among a set of $m$
initial variables measured on $n$ objects through a reduced number
$(p<m)$ of linearly independent factors $F1,F2 ... Fp$. 
The final aim of MFA is to reduce the number of independent variables
(factors) required to account for the variance of a class of objects.
This method has been used in astronomy by several authors (e.g.
Whitemore 1984; Vader 1986; Patat et al. 1994; Petrosian \&
Turatto 1992, 1995). A detailed description of MFA method can be
found in Harman (1967) and Afifi \& Azen (1979).

The initial variables used for the isolated parent galaxies of SNe
were the morphological type - $T$, the absolute magnitude of the
SN parents - $M_B$, the inclination of parent galaxy - $INC$, a
dummy parameter - AGN for the nuclear activity (dAGN=0 for normal
and 1 for active nuclei), the logarithmic ratio -
Log$L_{FIR}/L_B$, a dummy parameter - SN, for the SN type (dSN=1
for SNIa, 2 for core-collapse SNe, 1.5 for unclassified SNe) and
the deprojected relative distances of SNe from the centers of
parent galaxies - $R_0/R_{25}$.

The accumulated dispersion carried by the first three factors is
about 66\%. Adopting a correlation threshold of 0.7 we find that
the first factor, $F1$,  correlates the morphological type of SN
host galaxies with the logarithmic ratio Log$L_{FIR}/L_B$, with
early type galaxies having smaller IR excess, a well known finding
for all galaxies.  Factor two, $F2$, correlates the relative
distance $R_0/R_{25}$  of SNe and the absolute blue magnitude of
the host galaxies with brighter galaxies hosting the more distant
SNe from the nucleus. This is most likely a selection effect of SN
searches. In magnitude limited samples, more distant galaxies are
on average intrinsically brighter  than nearby galaxies; at the
same time,  because of the Shaw effect, SNe in distant galaxies
are discovered at higher distance from the nucleus (e.g.
Cappellaro \& Turatto 1997). The third factor, $F3$, accounts
mainly for the SN type.

According to the theory of MFA the factors are orthogonal, hence
the corresponding initial variables are independent. It means that
for the isolated galaxy sample there is no relation between SN
type, their radial distribution and the parameters of the host galaxy,
in particular their FIR luminosity excess, which is thought to 
characterize their star formation activity.

If we exclude from the analysis the galaxy inclination, thus
extending the sample also to non-disk dominated galaxies, only two
factors have eigenvalues larger than 1. The first factor, $F1$,
correlates the nuclear activity type of SN host galaxies with the
Log$L_{FIR}/L_B$ with also a significant contribution from the
morphological type. The second and third factors remain
unaffected.

A similar factor analysis was conducted for isolated pairs using
the same parameters described above, plus the logarithmic ratio of
the blue luminosity of SN parent to that of the neighbor galaxy -
Log$L_P/L_N$, the radial velocity differences - $\Delta V$,  the
ratio of the projected distance of SN from the center of parent
galaxy to the projected distance between the two members of the
systems - $R_{SN}/R_{1,2}$, and the dummy variable - PA,
indicating the location of SNe with respect to the companion (0
when the SN is on the companion side, 1 otherwise).

The accumulated dispersion carried by the first four factors is
about 68\%. With the correlation threshold of 0.7 the first
factor, $F1$, is the combination of morphology, activity class and
log$L_{FIR}/L_B$ , and can be understood considering that early
type galaxies host more frequently AGN and have smaller infrared
excess than late type galaxies.  $F2$ accounts mainly for 
$R_0/R_{1,2}$. $F3$ indicates a correlation between absolute blue 
magnitude and Log$L_P/L_N$, which results from the well known 
properties that SN rate per unit galaxy luminosity is a costant. 
$F4$ accounts mainly for $\Delta V$.

For the sample of  SNe in groups we have added two variables
characterizing the groups, i.e. the number galaxies in the group -
NUMGAL, the standard deviation of the heliocentric radial
velocities of the group members - SIGMA, and two variables
characterizing the SN host galaxy in the context of the group, i.e
the logarithmic ratio of its blue  luminosity to that of the most
luminous galaxy of the group - Log$L_P/L_{LG}$, and the distance
of the host galaxy from geometrical center of the group relative
to that of the most distant member of group - $R_P/R_{GR}$. In
this analysis the dummy variable  PA represents the SN location in
the direction -1, or in the opposite direction - 0, of the
geometrical center of the group.

The accumulated dispersion for the first four factors is rather
low, 60\%. F1 combines two parameters of the group NUMGAL and
SIGMA, with richer groups having higher velocity dispersion. $F2$
correlates the absolute magnitudes of SNe parents, $M_B$, and
Log$L_P/L_{LG}$, due to the presence in both variables of the
galaxy luminosity. $F3$ accounts mostly for the morphological type
of  SNe hosts. No clear correlation appears from $F4$.

Removing the galaxy inclination, i.e. considering all galaxy
types, the first three factors remain unchanged while $F4$
accounts only for PA.

Initial variables of orthogonal factors are independent, which
means that in groups of galaxies there is no relation between
type, location and distribution of SNe, total star formation
activity in galaxies of a given type and the parameters of the groups.

A more general exploratory analysis has been performed by merging
the three samples, considering the variables in common plus a
dummy parameter describing the multiplicity of the parent system
(1 for isolated, 2 for pairs and 3 for groups of galaxies). Three
first factors loadings accumulated about 55\% of the dispersion.
$F1$, is the morphological type of the SN host galaxies. $F2$
correlates absolute blue luminosity of SNe hosts with AGN class of
galaxies. $F3$ is related to the dummy parameter describing
different systems. It means that there is no correlation among the
parameters characterizing SNe, their host galaxies and different
systems of galaxies.

\section{Conclusions}

In order to investigate the influence of the environment on the
star formation we have used as indicators the SNe exploded in
isolated galaxies, in pairs and in groups of galaxies. Indeed,
core collapse SNe which descend from massive, short-lived
progenitors, explode only few millions years after the births of
the progenitors and can be considered as probes of  recent star
formation activity (Cappellaro et al. 1999).

The samples, described in Sect. 2, were sufficiently large to
derive SN rates, radial distributions inside the parent galaxies,
and for SNe in multiple systems, also the locations inside the
systems.

The mean morphological types and luminosities of the SN host
galaxies in different systems are similar. Contrary to the case of
core collapse SNe, the average absolute magnitudes of parent
galaxies of SNIa is brighter in paired galaxies than in other
systems. Currently, we have no interpretation for this finding.

We found that the rate of SNe in galaxies member of pairs is 
$\sim$40\% higher than in  the average galaxy and $\sim 60$ higher 
than in galaxies member of groups. 
Although, due to the large uncertainties this is not a solid 
conclusion, we can speculate that the higher SN rate is consistent 
with the enhanced star formation rate in close interacting system 
and instead more mild interaction, as that occuring in groups, 
does not affect significantly the SF rate. We remark that since only 
a fraction of the galaxy pairs experience strong
interaction, the effect on the complete sample is expected to be diluted.

Also the radial distributions of core-collapse SNe do not differ
in various systems and are comparable to the radial distributions
founds by Bartunov et al. (1992) on a general sample. The
azimuthal distributions of SNe inside their parent galaxies in
relation to the location of the neighbors in pairs and groups of
galaxies does not reveal any effect. It must be noted that,
because of observational limitations, our investigation does not
probe the SN frequencies and distributions in the circumnuclear
regions of the galaxies.

An extensive Multivariate Factor Analysis has been performed
including parameters of the  SNe, physical and geometrical
parameters of the parent galaxies and of the environment. A number
of known and/or obvious correlations have been pointed out.
Anyway, no clear correlation has been revealed between the
environment of the host galaxies and the hosted SNe. 

We conclude, therefore, that there is no clear effect of the galaxy
environment on the SN production, with the possible exception of strongly 
interacting systems,  and that the properties of the host 
galaxies are the same in field and in denser surroundings. 

\section*{Acknowledgments}
H.N. And A.R.P. acknowledges the hospitality of the astronomical
Observatories of Padova (Italy) and Marseille (France). H.N. Work
in Marseille Observatory was supported by the grant of French
Government. A.R.P. Work in Padova Observatory was supported by
NATO grant PST.CLG 977376. This research has made use of the
Lyon-Meudon Extragalactic Database (LEDA) supplied by the LEDA
team at the CRAL-Observatoire de Lyon and NASA/IPAC Extragalactic
Database (NED), which is operated by the Jet Propulsion
Laboratory, California Institute of Technology, under contract
with the National Aeronautics and Space Administration.

We also aknowledge the helpful comments of an anonymous referee.

\end{document}